\documentclass[lettersize,journal]{IEEEtran}
\usepackage{amsmath,amssymb,amsfonts}
\usepackage{algorithmic}
\usepackage{algorithm}
\usepackage{array}
\usepackage[caption=false,font=normalsize,labelfont=sf,textfont=sf]{subfig}
\usepackage{textcomp}
\usepackage{stfloats}
\usepackage{url}
\usepackage{verbatim}
\usepackage{graphicx}
\usepackage{cite}
\usepackage{mathtools}
\usepackage{xcolor}

\begin{document}

\title{ISAC-Assisted Wireless Rechargeable Sensor Networks with Multiple Mobile Charging Vehicles}

\author{Muhammad Umar Farooq Qaisar, Weijie Yuan, Paolo Bellavista, Guangjie Han, and Adeel Ahmed
\thanks{Muhammad U. F. Q. and Weijie Y. are with the School of System Design and Intelligent Manufacturing, Southern University of Science and Technology, Shenzhen, China}
\thanks{Paolo B. (paolo.bellavista@unibo.it) is with the Department of Computer Science and Engineering, University of Bologna, Bologna, Italy}
\thanks{Guangjie H. (hanguangjie@gmail.com) is with the Department of Internet of Things Engineering, Hohai University, Changzhou, China}
\thanks{Adeel A. (adeelahmed@mail.ustc.edu.cn) is with the School of Computer Science and Technology, University of Science and Technology of China, Hefei, China}
\thanks{Corresponding authors: Weijie Yuan and Muhammad Umar Farooq Qaisar, Email:\{yuanwj, muhammad\}@sustech.edu.cn}}

\maketitle

\begin{abstract}
As IoT-based wireless sensor networks (WSNs) become more prevalent, the issue of energy shortages becomes more pressing. One potential solution is the use of wireless power transfer (WPT) technology, which is the key to building a new shape of wireless rechargeable sensor networks (WRSNs). However, efficient charging and scheduling are critical for WRSNs to function properly. Motivated by the fact that probabilistic techniques can help enhance the effectiveness of charging scheduling for WRSNs, this article addresses the aforementioned issue and proposes a novel ISAC-assisted WRSN protocol. In particular, our proposed protocol considers several factors to balance the charging load on each mobile charging vehicle (MCV), uses an efficient charging factor strategy to partially charge network devices, and employs the ISAC concept to reduce the traveling cost of each MCV and prevent charging conflicts. Simulation results demonstrate that this protocol outperforms other classic, cutting-edge protocols in multiple areas.
\end{abstract}


\section{Introduction}
\begin{figure*}[h]
    \centering
	\includegraphics[width=\linewidth,height=8cm]{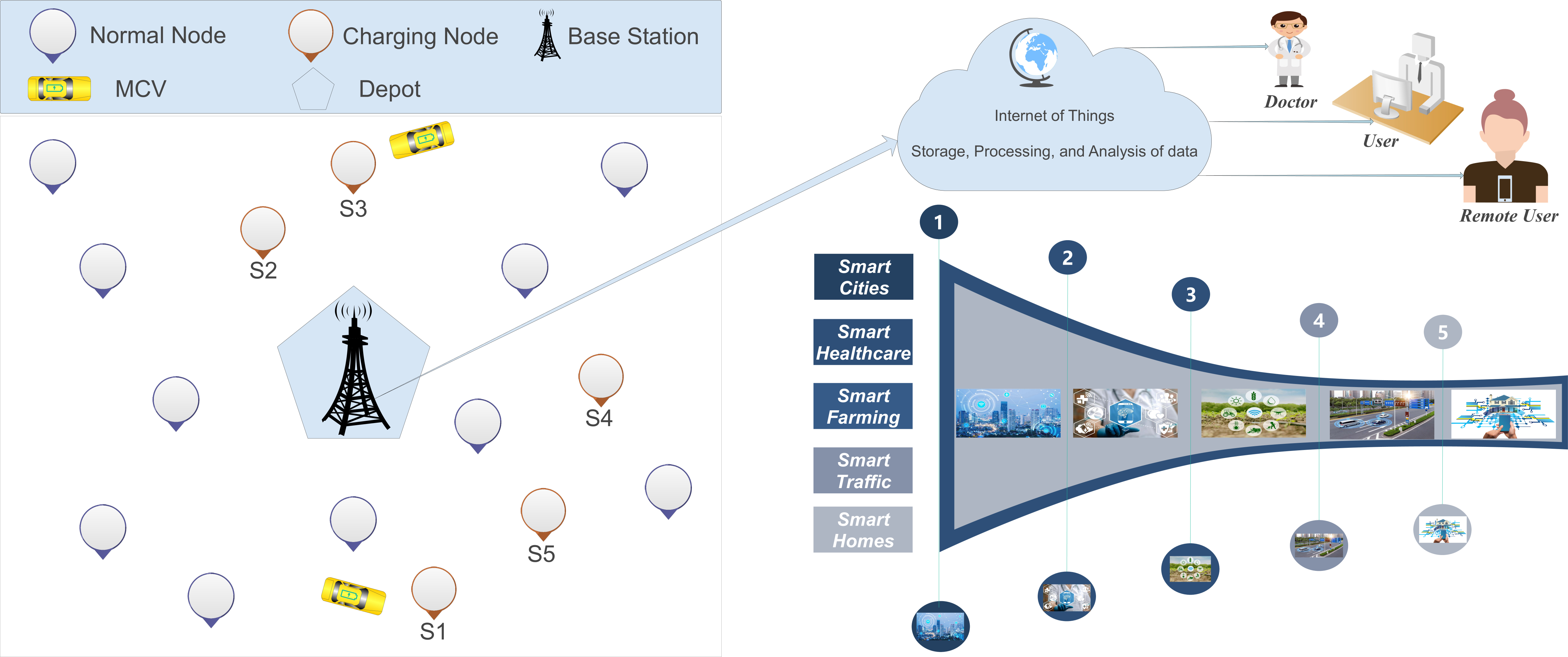}
	\caption{Wireless rechargeable sensor networks for IoT applications}
	\label{fig:WRSN}
\end{figure*}

The Internet of Things (IoT) has revolutionized the way we interact with technology, from smart homes to wearable devices \cite{1R}. Central to this transformation are wireless sensor networks (WSNs), which enable the connection and information transmission of devices and systems. However, WSNs face a significant challenge in the form of an energy shortage, which can impact their capability to function effectively \cite{2}. This is where the concept of wireless rechargeable sensor networks (WRSNs) comes in, utilizing wireless power transfer (WPT) technology to ensure energy sustainability. WPT is employed to wirelessly recharge the energy-starved sensor devices. Three popular WPT technologies are inductive coupling, electromagnetic (EM) radiation, and magnetic resonant coupling. In contrast to the initial two techniques, the magnetic resonant coupling exhibits superior energy transfer efficiency under omni-direction, eliminates the need for line-of-sight (LOS), and is unaffected by external factors. WRSNs have practical applications in various fields of IoT, including smart cities, smart healthcare, smart farming, smart traffic, and smart homes. Typically, a WRSN is comprised of a base station, which serves as a depot for mobile charging vehicles (MCVs), one or more MCVs, and sensor devices equipped with rechargeable batteries that receive wireless signals to recharge from MCVs as depicted in Figure \ref{fig:WRSN}. This unique charging approach enables WRSNs to operate efficiently and continuously, without interruption \cite{3R}.

To recharge the sensor devices in WRSNs, the MCVs use either periodic charging or on-demand charging strategies. While periodic charging follows a predetermined schedule, it is not always ideal due to the dynamic energy depletion rate of the sensor devices. In contrast, on-demand charging is more flexible and is capable of making real-time decisions based on the energy requirements of the sensor devices. Additionally, charging strategies can be either full or partial charging models. Full charging results in significant charging delays, whereas partial charging allows for more sensor devices to be recharged. Nevertheless, existing strategy designs fail to take into account the traveling time and potential conflicts between multiple MCVs, which are also of great importance in the charging process \cite{4}.

We intend to address the aforementioned problem by utilizing the innovative Integrated Sensing and Communication (ISAC) technique \cite{5}. By integrating the functionalities of sensing and communications, this technique allows for efficient utilization of wireless resources, wide-area environmental sensing, and mutual benefits. Moreover, by exploiting the benefits of wireless signals \cite{6}, ISAC can improve the charging efficiency of MCVs and reduce their travel time. In this work, the charging sensor devices can be arranged in a queue, with multiple MCVs having different priorities. When an MCV comes within range of a prioritized charging sensor device, the sensor device transmits an ISAC signal to the vehicle. After receiving an echo of the signal via wireless transmission, the sensor device analyzes it and communicates with the base station to update the charging priorities of other MCVs in the queue. This approach improves the efficiency of charging by minimizing travel time and avoiding conflicts that might arise when multiple MCVs attempt to charge the same sensor device.

The studies mentioned above have identified several compelling reasons for addressing the challenges associated with deploying multiple MCVs and developing an effective on-demand charging strategy for sensor devices, as well as integrating the ISAC concept with WRSNs to optimize network stability. Thus, this article introduces the ISAC-assisted WRSNs protocol, which includes three key strategies.  The first strategy ensures a balanced charging load across each MCV priority queue by taking into account four attributes: residual energy of the charging device, the distance between the MCV and the charging device, the degree of the charging device, and the charging device betweenness centrality. The second strategy determines the charging factor for each MCV queue, enabling the partial charging of all sensor devices to further enhance charging efficiency and network lifetime. Finally, the third strategy integrates the ISAC concept with WRSNs to leverage wireless resources, minimize travel time, and avoid conflicts that may occur when multiple MCVs seek to charge the same sensor device. The main goal of this study is to propose a novel charging strategy for establishing a balanced load distribution across several MCVs using a highly effective on-demand charging technique. This strategy will result in noticeable improvements in the overall effectiveness of charging. In addition, the study presents a cutting-edge sensing and communication technique that significantly reduces the amount of time MCVs spend within the network. The findings of our developed protocol show that it outperforms more recent state-of-the-art protocols in terms of performance and provides strong evidence of its ability to improve MCV charging efficiency while decreasing travel time.

\section{Related Work}
The significance of energy replenishment in WRSNs grows as we progress deeper into the IoT domain. This section briefly examines the studies on WRSN energy replenishment that are pertinent to our work and sheds light on the most recent developments and cutting-edge approaches in this field.

The authors of \cite{7} state in a paper that they have developed a charging method that clusters each device's energy requirements in order to provide an equal distribution of the charging load across MCVs. This technique significantly increases the number of recharged devices while decreasing charging time. To reduce charging delays, a charging scheduling mechanism is presented in \cite{9}. The authors set up a closed charging tour for each MCV in an effort to stop sensor devices from being charged by multiple MCVs at once. Their method, meanwhile, led to an uneven distribution of charging loads among the MCVs. A distributed mobile charging methodology is proposed in \cite{10} and is intended to schedule multiple MCVs in congested WRSNs. The authors utilized game theory techniques to tackle the issue of multi-charging and employed an on-demand partial charging strategy that resulted in a repetitive game played by the MCVs. The authors claimed that their method resulted in better charging coverage and a decrease in charging time. In \cite{11}, the authors tackled the challenge of coordinating multiple MCVs to schedule and optimize charging with the aim of reducing the overall energy consumption of the MCVs. Their approach involved modifying the mobility speed and charging time to minimize travel time and improve efficiency. The study presented by the authors in \cite{12} is an uneven cluster-based mobile charging method that divides sensor devices into groups and optimizes the charging schedule for each MCV based on residual energy and distance to sensor devices. However, their method leads to reduced charging efficiency. A study in \cite{13} proposed a charging scheduling approach that uses fuzzy logic to manage multiple MCVs. The authors aimed to equally share the charging load of sensor devices among the MCVs by dividing the network. They also established dynamic charging thresholds for each sensor device based on its rate of energy consumption. Additionally, the authors used fuzzy logic and multi-metric inputs to determine the next sensor device to be charged for each MCV. Their approach falls short in efficiently selecting the next sensor device to be charged due to the inefficiency of the multi-metric strategy employed. A recent work in \cite{14} developed a novel approach to minimize charging delays in WRSNs through charging sensor devices with multiple MCVs. This method, in contrast to earlier works, used predetermined MCV travel trajectories with varying speeds.

Previous research on charging scheduling strategies in WRSNs has demonstrated their feasibility and potential usefulness. These approaches, however, lacked effectiveness in balancing charging loads among MCVs and implementing a viable charging factor strategy for partially charging network devices. Furthermore, they were inadequate in addressing the problem of minimizing travel time for multiple MCVs and resolving charging conflicts in the network. To address these issues, this article develops a novel scheme that employs an effective multi-metric and charging factor strategy while also introducing the concept of ISAC to WRSNs. This concept optimizes the sensing and communication tasks between charging devices and MCVs, resulting in reduced travel time and charging conflicts. This approach goes beyond the current state-of-the-art WRSN charging scheduling strategies.

\section{System Model}

\begin{figure*}[t]
    \centering
	\includegraphics[width=\linewidth,height=8cm]{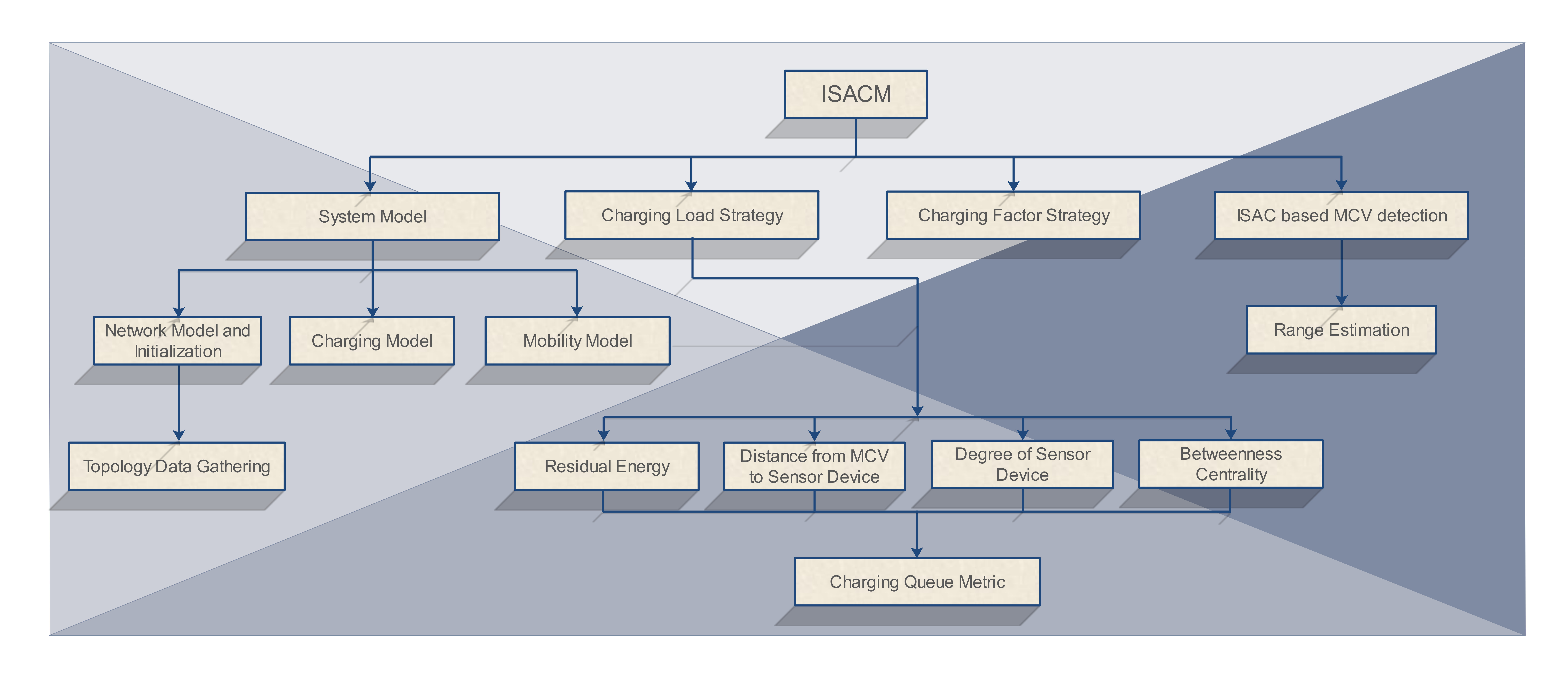}
	\caption{The core architecture of the proposed novel scheme}
	\label{fig:CAPP}
\end{figure*}

\subsection{Network Model and Initialization} \label{NMA}
With the assumption of a wireless rechargeable sensor network, we have a set of randomly deployed sensor devices and a set of MCVs in a two-dimensional region. When the energy level of the sensor devices drops below a specific threshold, they send charging requests to the base station, which is positioned at the central location within the region and serves as a depot for the MCVs. This research adapts the initial position of the MCVs, described in \cite{10}, to determine the coordinates of the MCVs' initial position in $k$ different regions. At a time, the MCV recharges only one sensor device. Obviously, the energy capacity of an MCV is significantly higher than that of a sensor device.

\subsection{Charging Model}
According to the WPT technique, the sensor device can only receive a fraction of the power transmitted by the MCV. The WPT efficiency between an MCV and a sensor device is non-linearly correlated with the distance between the two, indicating that the charging efficiency decreases as the distance increases. The charging process is considered successful only when the MCV arrives at the precise location of the sensor device. Therefore, we aim to achieve the desired efficiency by employing a partial-charging strategy, and the requesting sensor devices may not be fully recharged. It is important to note that the effectiveness of WPT is consistent across all sensor devices. However, MCV has limited energy capacity and must return to the sink for self-recharging when its residual energy falls below a predetermined threshold. The base station serves as a depot for the MCVs.

\subsection{Mobility Model}
We adopt a recent and widely accepted extension of the Random Waypoint (RW) mobility model, which has been used in a recently published state-of-the-art article \cite{16R}. The charging time of the sensor devices is described by the pause value at each stopover location and the positioning plan based on the direction of the subsequent sensor devices in the queue awaiting charging. We assume that the MCV travels along a straight, obstacle-free path at a constant speed, and its location is continually updated until it reaches the location of the next sensor device in line to be charged. When the distance between the MCV and the charging sensor device is less than or equal to the distance threshold for efficient energy transmission, the charging process takes place.

\section{The Proposed Protocol}
In this article, we propose a technique for balancing the charging load on each MCV and partially charging all sensor devices in a queue in order to improve charging efficiency. We also incorporate the ISAC technique to make use of wireless resources to reduce the network travel time of the MCVs. This strategy can be applied to potential IoT applications where network coverage and energy efficiency are important considerations. Figure \ref{fig:CAPP} displays the basic design of the proposed \textit{ISACM}.

\subsection{Charging Load Strategy} \label{CLS}
In this section, we present a strategy for balancing the charging load for each MCV queue in the wireless rechargeable sensor network. To accomplish this, we consider four important attributes: residual energy of the charging sensor device, distance from MCV to charging sensor device, degree of a charging sensor device, and charging sensor device betweenness centrality. Each attribute is associated with a probability distribution function, and the charging queue metric is determined by taking their respective values into account. The charging sensor devices with the highest value of this metric are prioritized in the charging queue. Because the MCVs are distributed in $k$ different locations throughout the network, each MCV will have a distinct sequence of prioritized charging sensor devices.

\begin{figure*}[b]
    \centering
	\includegraphics[width=0.8\linewidth,height=8cm]{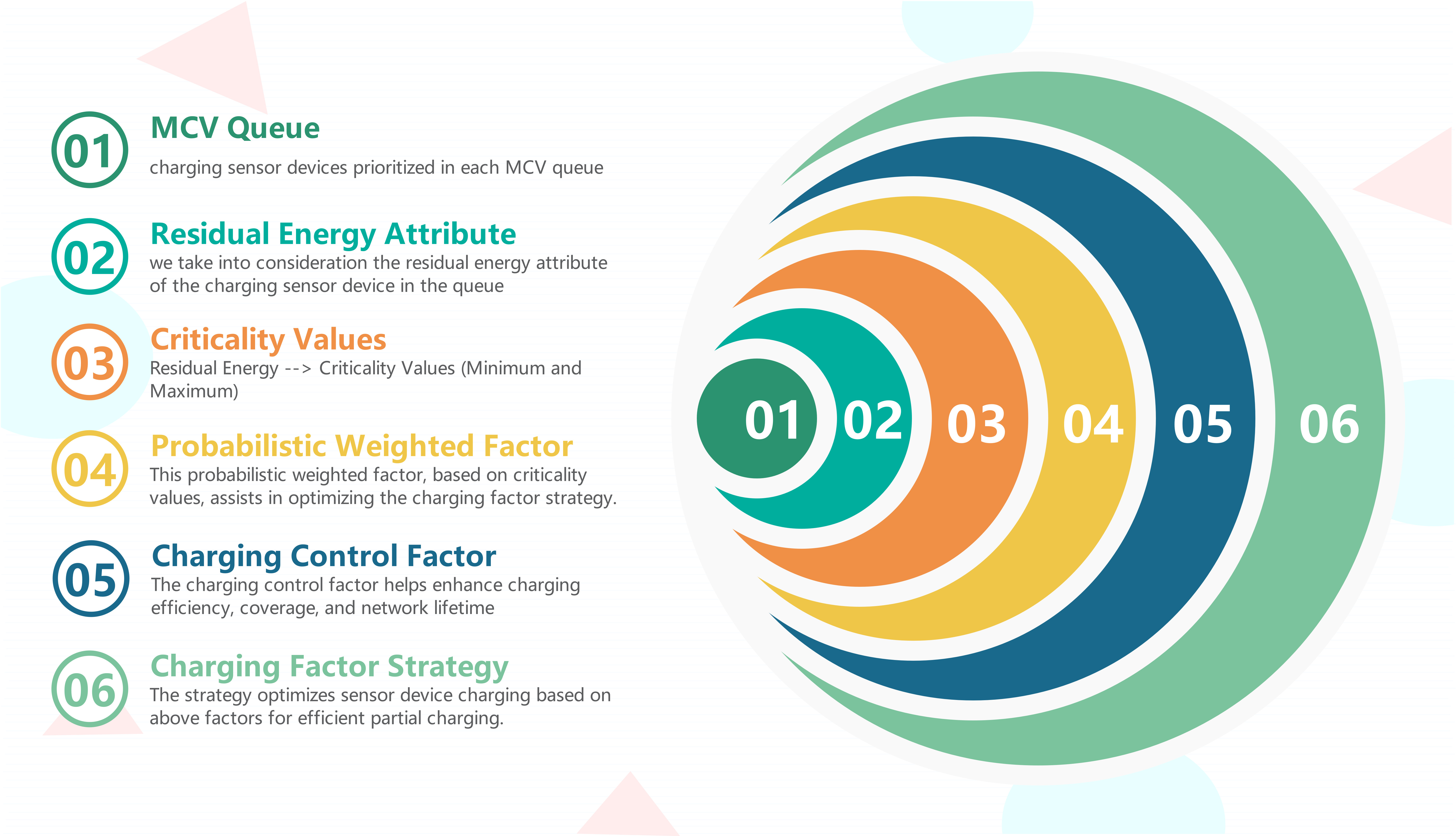}
	\caption{The figure illustrates the steps involved in the charging factor strategy: 1) It starts with the charging sensors in the MCV queue. 2) The residual energy attribute of the charging sensor device is taken into consideration in the queue. 3) The residual energy values are then used to calculate the criticality values, representing the minimum and maximum criticality of the sensor devices. 4) These criticality values are then utilized in the calculation of the probabilistic weighted factor. 5) The charging control factor is obtained for each sensor device in the queue to regulate the charging process efficiently, adjusted to 10\% of the residual energy priority for each sensor device. 6) Finally, the charging factor strategy is determined based on the above factors to enhance charging efficiency and ensure fair and successful partial charging.}
	\label{fig:CFS}
\end{figure*}

\subsubsection{Residual Energy of Charging Sensor Device} \label{RECN}
In the context of IoT networks, optimal scheduling of charging sensor devices is essential with residual energy being an important factor to consider. This section explains how to efficiently use residual energy as a scheduling variable to prioritize the charging process. The attribute reflects a charging sensor device's residual energy below the residual energy threshold and prioritizes sensor devices with lower residual energy than other charging devices.

\subsubsection{Distance from MCV to Charging Sensor Device} \label{DMCN}
In IoT networks, it is essential to effectively schedule charging sensor devices, and the distance between MCV and charging sensor devices must be taken into account. Sensor devices will be prioritized according to how far they are from the MCV with the help of this attribute. The devices with the shortest distance will be given the highest level of priority. 

\subsubsection{Degree of Charging Sensor Device} \label{DCN}
The degree of a charging sensor device in an IoT network, which can be estimated by the number of connections it has with adjacent devices, is essential to the data flow rate of that device. 
A higher number of neighbors results in a greater probability of higher data flow to the charging sensor device. Moreover, if a device has low energy levels below the charging threshold, a relatively high data flow rate from numerous neighbors could quickly deplete its residual energy, thereby making efficient scheduling highly demanded. Therefore, this attribute aims to prioritize sensor devices with the maximum number of neighbors among all the charging sensor devices.

\subsubsection{Charging Sensor Device Betweenness Centrality} \label{CNBC}
The concept of betweenness in an IoT network refers to the number of times a sensor device acts as a bridge between two other devices along the shortest path. As such, the more frequently a charging sensor device acts as a betweenness, the greater the chance of information flow, which can simultaneously deplete its energy and make it critical. Therefore, the purpose of this attribute is to prioritize charging sensor devices that frequently serve as bridges.

The attributes are normalized within the range of 0 and 1. Based on this normalization, we determine the probability distribution function of each attribute through curve fitting. Additionally, we introduce a weighted factor for each attribute to augment the influence of the probability distribution functions and prioritize charging sensor devices by increasing their respective values.

\subsubsection{Charging Queue Metric}
The objective of the charging queue metric is to effectively prioritize the charging queue of each MCV by utilizing the four aforementioned attributes. As outlined in Section \ref{NMA}, the initial location of each MCV is partitioned into $k$ regions. Therefore, depending on the distance attribute between the MCV and the sensor device, the priority of charging sensor devices in each MCV queue will vary. In this work, each MCV queue's priority is determined by the base station using the charging queue metric. The average value of all four attributes is used to calculate the charging queue metric, which is given the highest priority value.

\subsection{Charging Factor Strategy} \label{CFS}
The probabilistic partial charging model presented in this section focuses on the charging factor strategy to increase charging efficiency by taking into account the criticality of charging sensor devices based on their residual energy attributes. The probabilistic weighted factor plays an vital role in determining the order in which charging tasks should be completed, and the presence of the charging control factor ensures fair and effective partial charging, improving charging efficiency, coverage, and longevity of the network. The charging factor strategy is depicted in detail in Figure \ref{fig:CFS}.  

\subsection{ISAC-based MCV detection}
In the context of WRSN, the travel costs incurred by MCVs are directly related to ISAC. To increase the overall efficiency of WRSNs, ISAC implies the mutual improvement of sensing and communication capabilities. When it comes to MCV travel costs, ISAC plays an essential role in maintaining the balance between the relationship between travel costs, travel time, and rewards. To increase the energy usage efficacy of wirelessly charging sensors, ISAC attempts to decrease the cost of travel. Thus, the relationship between ISAC and MCV travel costs in WRSNs is based on optimizing energy usage for recharging and data collection, which are critical to the network's optimal operation.
 
\begin{figure}[h]
    \centering
	\includegraphics[width=0.8\linewidth,height=4cm]{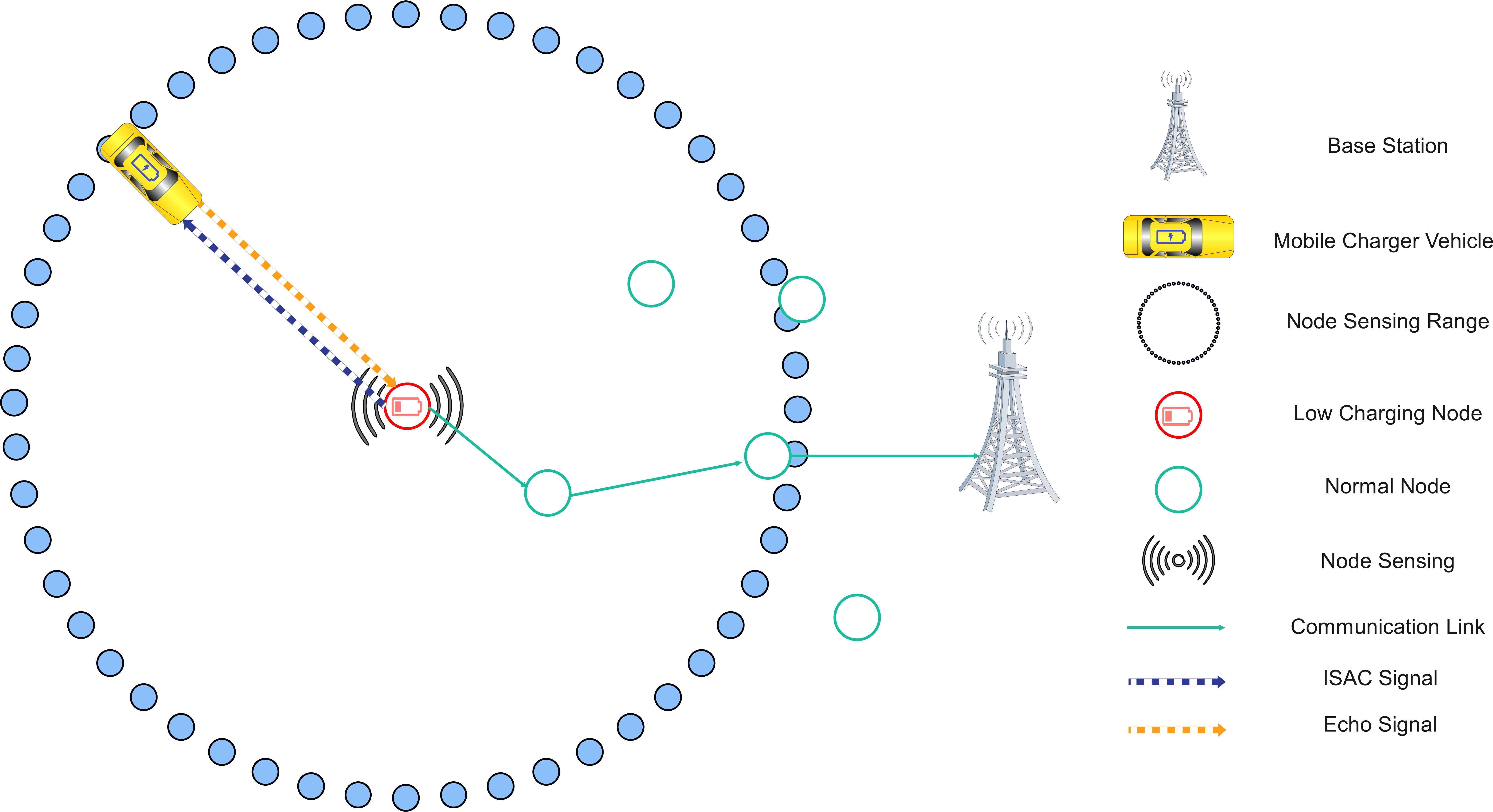}
	\caption{MCV detection based on ISAC approach}
	\label{fig:ISACMCV}
\end{figure}

In the ISAC approach, the charging sensor device in the prioritized queue transmits an ISAC signal to the closest MCV within its sensing range. With the use of the ISAC signal, the device can determine distance by analyzing the received echo signal while accounting for noise and interference. Figure \ref{fig:ISACMCV} illustrates how the device then engages in communication with the base station to avoid other MCVs from approaching and overcharging it. 

This cuts down on the time it takes for MCVs to charge and their travel time. By evaluating the information collected using this method, it enables the base station to modify the priority of other MCVs' charging queues.

The objective is to extract information from a cluttered received signal because both noise and interference affect the received echo signal. In order to do this, the sensor device applies matched filtering to the signal it receives. The matched filter is specifically designed to increase the correlation between the received signal and the signal sent by the MCV.

To achieve the maximum cross-correlation function time delay, we begin by rewriting the cross-correlation function as a function of time delay and then compute the derivative of the cross-correlation function with respect to time delay. The optimal time delay that maximizes the cross-correlation function is the one that maximizes the convolution of the received echo signal and conjugate of the signal transmitted by the MCV.

The charging sensor device estimates the distance to the MCV by maximizing the cross-correlation function with a time delay and using the speed of light to calculate the distance. If the distance is within the sensing region, the sensor device detects the MCV and communicates with the base station according to the probabilistic sensing model \cite{15}. Following that, the base station removes that sensor device and updates the priority queues of other MCVs.

\section{Performance Evaluation and Discussions}
In order to evaluate the performance of the proposed protocol, a carefully designed simulation setup was employed to conduct comprehensive simulations. The WRSN network is set up in a square monitoring area, with sensor devices distributed at random. The base station, which is positioned in the middle of the area, acts as a depot where MCVs may replenish their batteries. The scheduling of charging requests for each MCV and network management are also its responsibilities. We performed 20 random simulations, averaged the results, and verified the accuracy of our results. Table \ref{SP} presents a summary of the important simulation parameters and their corresponding values.

\begin{table}[h]
	\centering
	\caption{Simulation Parameters}
	\label{SP}
	\begin{tabular}{p{4cm}p{3.5cm}}
		\hline
		\textbf{Parameter} & \textbf{Value} \\
		\hline
		Number of sensor devices  & Varies from $100$ to $500$\\
		Communication range & $50m$ \\
  	Sensing range & $25m$ \\
		Sensor device battery capacity & $0.5J$ \\
        Threshold for charging requests & $30\%$ residual energy \\
		MCV battery capacity & $10kJ$ \\
  	The charging rate & $0.05J/s$ \\
		MCV travel speed & $5m/s$ \\
		MCV travel cost & $5J/m$ \\
		\hline
	\end{tabular}
\end{table}

\begin{figure*}[t]
\centering
\subfloat[\scriptsize Energy Usage Efficiency]{\label{fig: EUENN}\includegraphics[width=.32\linewidth, height=5cm]{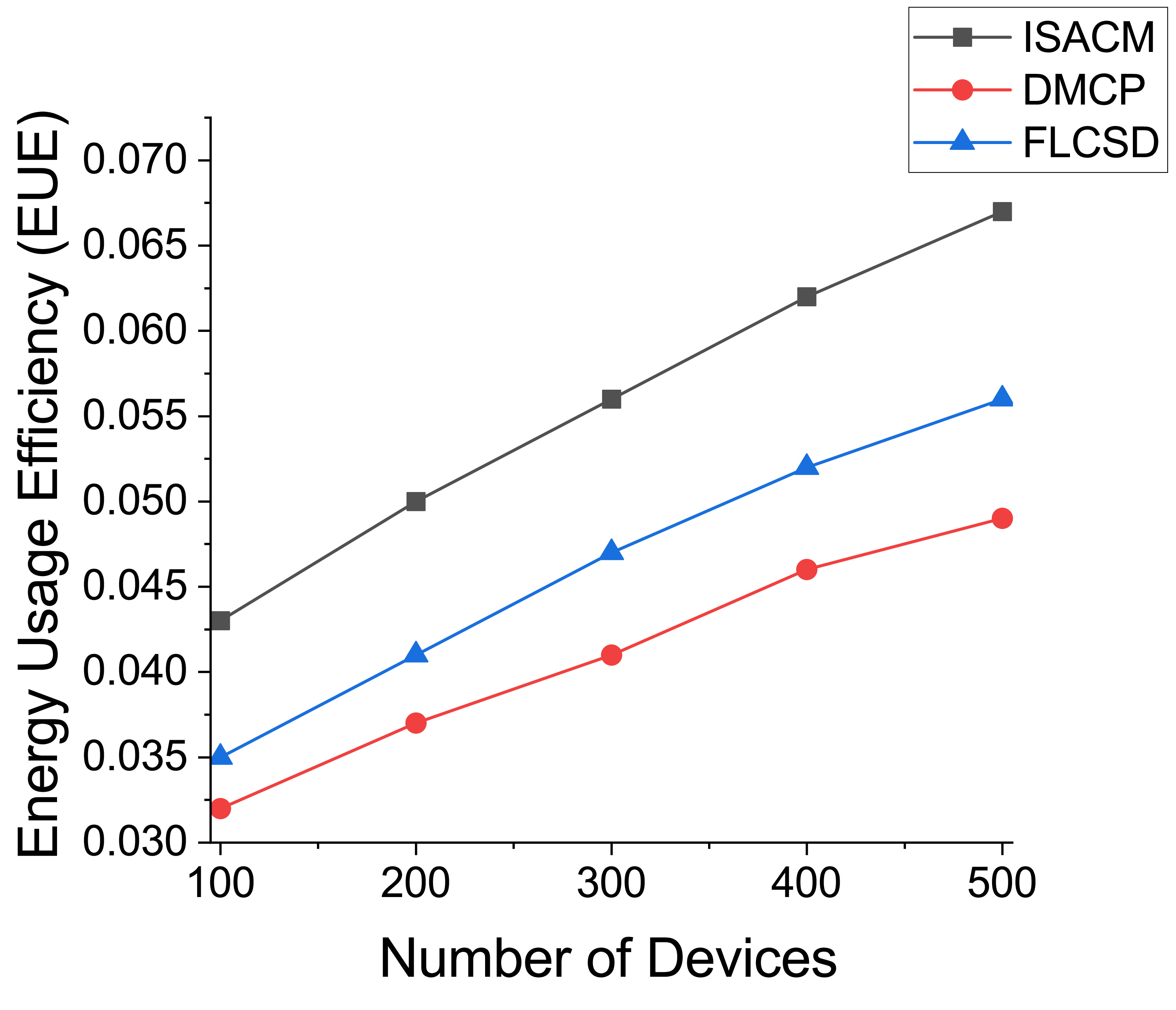}}\quad
\subfloat[\scriptsize Charging Delay]{\label{fig: CDNN}\includegraphics[width=.32\linewidth, height=5cm]{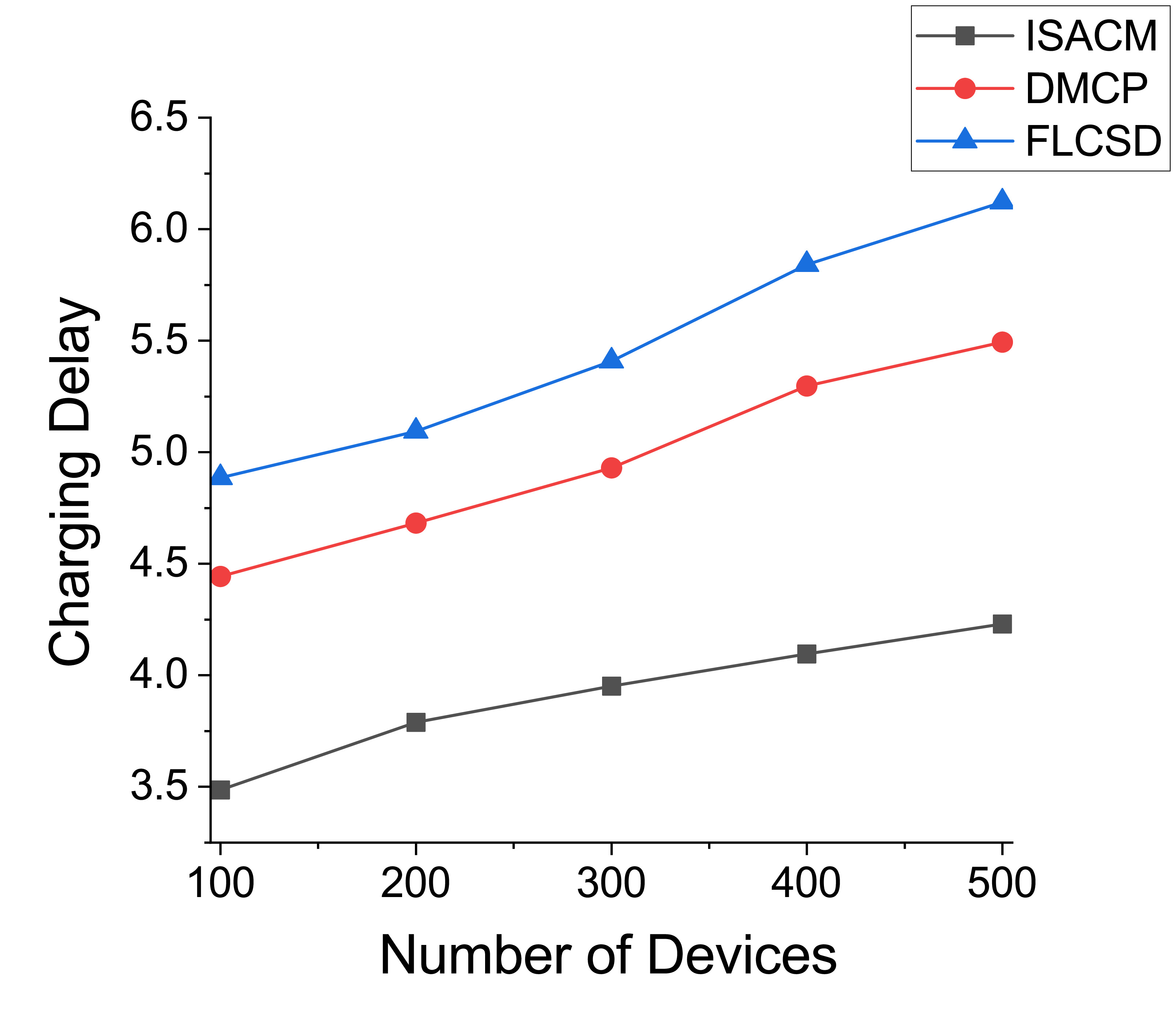}}\quad
\subfloat[\scriptsize Travel Distance]{\label{fig: TDNN}\includegraphics[width=.32\linewidth, height=5cm]{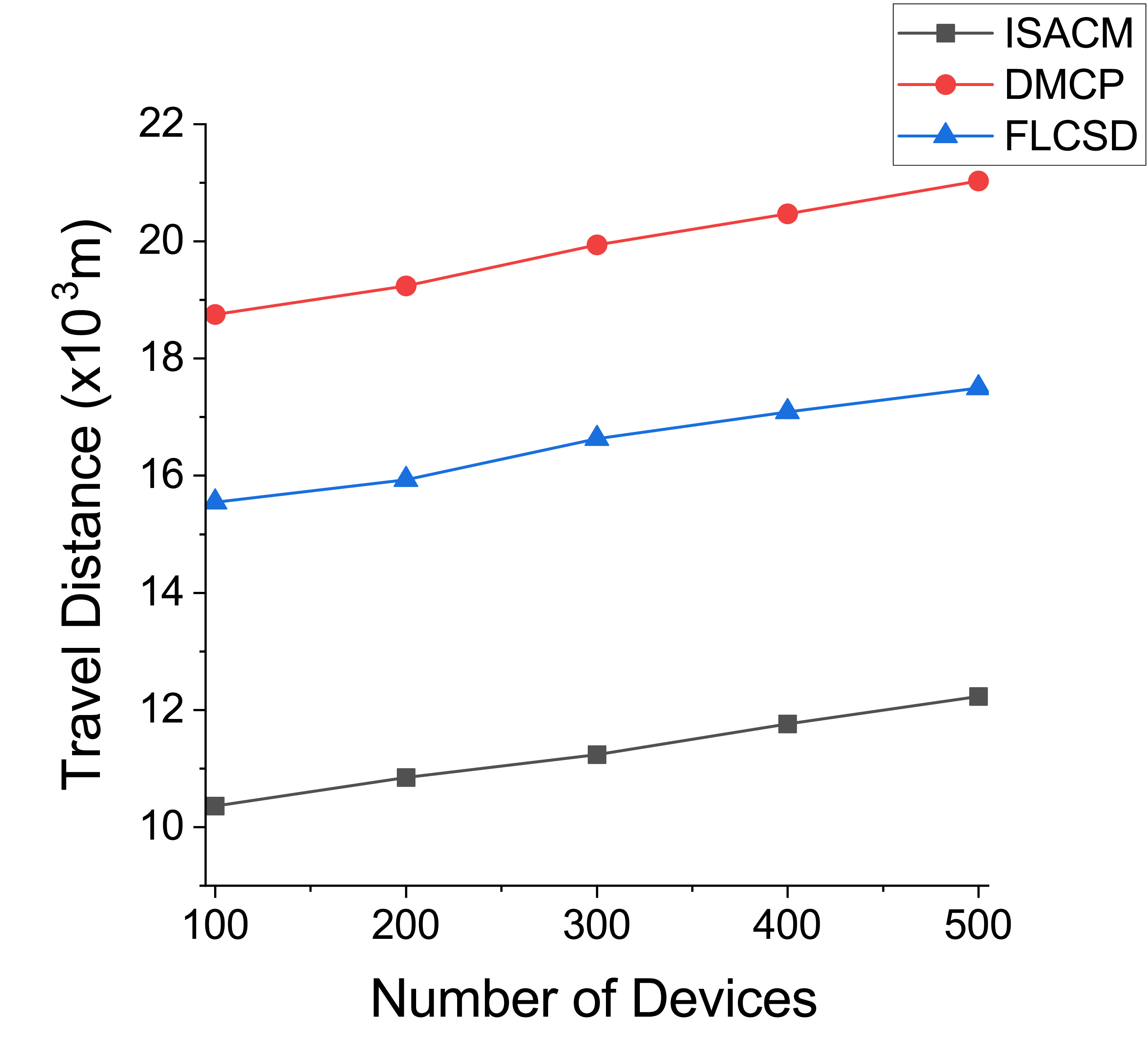}}
\caption{Performance over number of devices}
\label{fig: PNN}
\end{figure*}

The performance of the proposed protocol is measured based on several key parameters. These parameters include:

i) \textbf{Energy Usage Efficiency}: This parameter is defined as the ratio of the total energy transferred to the sensor devices to the total energy transmitted from the base station to the MCVs.

ii) \textbf{Charging Delay}: This parameter is defined as the time it takes for the MCVs to fulfill the energy requirements of the sensor devices.

iii) \textbf{Travel Distance}: This parameter is defined as the total distance covered by the MCV during a single charging tour.

The proposed protocol \textit{(ISACM)} is compared with two recent and state-of-the-art protocols, namely DMCP \cite{10} and FLCSD \cite{13}, to assess its effectiveness.

The energy usage efficiency results for the proposed protocol, FLCSD, and DMCP are depicted in Figure \ref{fig: EUENN}. The proposed protocol outperforms the state-of-the-art protocols due to several factors. Firstly, it employs a charging load strategy that prioritizes charging sensor devices in each MCV queue fairly. The strategy is based on a probability distribution, ensuring a balanced allocation of charging load. Secondly, an effective charging factor strategy is employed to partially charge all the sensor devices in the network. This strategy is based on the residual energy of the charging sensor device attribute. Finally, it maximizes the energy transferred to each requested sensor device in the network by employing the ISAC concept, which reduces the travel cost of each MCV. However, both FLCSD and DMCP protocols lacked efficient charging loads for each MCV in the network and prioritized charging queues, resulting in suboptimal energy usage efficiency.

The charging delay results are depicted in Figure \ref{fig: CDNN}, showing a gradual increase in delay with the number of sensor devices for all protocols. The proposed protocol achieves better performance compared to existing protocols by utilizing a probabilistic partial charging approach that covers more sensor devices in each MCV queue. In this method, a charging factor is assigned to each sensor device in the MCV queue based on its level of criticality, which assists in effectively partially charging the sensor devices. The strategy for the charging factor involves using the charging control factor, which aims to optimize the accessibility of each sensor device in the queue. Furthermore, the protocol adopts the ISAC concept to minimize travel costs and enhance the likelihood of charging the intended sensor device in an efficient manner. FLCSD protocol did not incorporate the partial charging approach. In contrast, DMCP protocol considered only the relative criticality and did not utilize a charging factor strategy that involves a probabilistic approach and charging control factor to decrease charging delay.

Figure \ref{fig: TDNN} illustrates the variation in travel distance with an increase in the number of sensor devices for all protocols. The proposed protocol exhibits superior performance compared to existing protocols as it prioritizes sensor devices in the charging queue that are closer to the MCV using the distance from the MCV to the charging sensor device attribute. Additionally, it adopts the ISAC approach as described in the charging delay result. In contrast, FLCSD and DMCP protocols did not consider minimizing the travel distance of the MCV in the network.

\section{Conclusion}
The presented work proposes an efficient charging solution for IoT applications through the use of ISAC-Assisted WRSNs with Multiple MCVs. The work centers around three key areas: load balancing charging scheduling for each MCV, an efficient charging factor strategy for partial charging of network devices, and an integrated sensing and communication approach to reduce the travel costs of MCVs. To accomplish this objective, the work initially evaluates four attributes with their corresponding probability distribution functions to balance the charging load and prioritize crucial sensor devices for charging. These attributes are the residual energy of the charging sensor device, the distance from MCV to the charging sensor device, the degree of charging sensor device, and the charging sensor device betweenness centrality. Next, it introduces an efficient strategy for charging that is based on the probability distribution function of residual energy of the charging sensor devices. Additionally, a charging control factor is included to minimize charging delays while increasing charging coverage. Lastly, the article utilizes the ISAC concept to identify the nearest MCV that arrives within the sensing range of the charging sensor device. This approach decreases the travel distance of other MCVs in the network and avoids charging conflicts between them. According to the simulation results, the proposed protocol exhibits superior performance compared to existing protocols.

\section{Future Directions and Challenges}
This section discusses potential future research directions as well as the current challenges in implementing ISAC-assisted WRSNs with multiple MCVs. It discusses the need for advancements in this area to improve the performance of WRSNs, as well as the open problems and challenges that still need to be addressed. This section aims to open the door for the development of effective charging strategies for IoT applications within WRSNs by addressing these future directions and challenges.

\subsection{Future Research Directions}
In the context of ISAC-assisted WRSNs with multiple MCVs, several promising research directions can be explored to further enhance the efficiency and effectiveness of the charging solution for IoT applications. First, the integration of machine learning algorithms holds great potential for optimizing the performance of such networks. By leveraging machine learning techniques, researchers can develop intelligent algorithms that adaptively control the charging operations, optimize resource allocation, and predict the charging demands based on historical data, network conditions, and user behavior patterns.

Another important research direction is the investigation of advanced energy harvesting techniques. Exploring novel methods such as solar, kinetic, or RF energy harvesting can significantly contribute to the self-sustainability of WRSNs by harnessing ambient energy sources. This would not only extend the lifetime of the sensor devices but also reduce the dependence on external power sources and increase the overall energy efficiency of the network.

\subsection{Open Problems and Challenges}
While ISAC-assisted WRSNs with multiple MCVs offer promising charging solutions for IoT applications, several open problems and challenges need to be addressed. One of the key challenges is energy optimization. Optimally managing and distributing the available energy resources among the sensor devices and MCVs to ensure uninterrupted and efficient charging is a complex task. Developing energy optimization algorithms that consider the dynamic nature of the network, the mobility of the MCVs, and the varying energy demands of the devices is crucial for maximizing energy utilization and network performance.

Scalability is another important challenge in deploying WRSNs with multiple MCVs. As the network expands to accommodate a larger number of sensor devices and MCVs, maintaining efficient communication, coordination, and resource allocation becomes increasingly complex. Designing scalable protocols and mechanisms that can handle the increased network size while minimizing overhead and preserving energy efficiency is essential for the widespread adoption of ISAC-assisted WRSNs. Additionally, cost-effectiveness, interoperability, and security also pose significant challenges that require further research and innovation to ensure the successful implementation and operation of these charging solutions in real-world IoT applications.

\bibliographystyle{IEEEtran}
{\footnotesize

\begin{thebibliography}{10}
\providecommand{\url}[1]{#1}
\csname url@samestyle\endcsname
\providecommand{\newblock}{\relax}
\providecommand{\bibinfo}[2]{#2}
\providecommand{\BIBentrySTDinterwordspacing}{\spaceskip=0pt\relax}
\providecommand{\BIBentryALTinterwordstretchfactor}{4}
\providecommand{\BIBentryALTinterwordspacing}{\spaceskip=\fontdimen2\font plus
\BIBentryALTinterwordstretchfactor\fontdimen3\font minus \fontdimen4\font\relax}
\providecommand{\BIBforeignlanguage}[2]{{%
\expandafter\ifx\csname l@#1\endcsname\relax
\typeout{** WARNING: IEEEtran.bst: No hyphenation pattern has been}%
\typeout{** loaded for the language `#1'. Using the pattern for}%
\typeout{** the default language instead.}%
\else
\language=\csname l@#1\endcsname
\fi
#2}}
\providecommand{\BIBdecl}{\relax}
\BIBdecl

\bibitem{1R}
A.~Khan, M.~M.~A. Khan, M.~A. Javeed, M.~U. Farooq, A.~Akram, and C.~Wang, ``Multilevel privacy controlling scheme to protect behavior pattern in smart iot environment,'' \emph{Wireless Communications and Mobile Computing}, vol. 2021, pp. 1--17, 2021.

\bibitem{2}
M.~U.~F. Qaisar, X.~Wang, A.~Hawbani, L.~Zhao, A.~Y. Al-Dubai, and O.~Busaileh, ``Sdorp: Sdn based opportunistic routing for asynchronous wireless sensor networks,'' \emph{IEEE Transactions on Mobile Computing}, 2022.

\bibitem{3R}
F.~T. Wedaj, A.~Hawbani, X.~Wang, M.~U.~F. Qaisar, W.~Othman, S.~H. Alsamhi, and L.~Zhao, ``Reco: On-demand recharging and data collection for wireless rechargeable sensor networks,'' \emph{IEEE Transactions on Green Communications and Networking}, 2023.

\bibitem{4}
A.~Kaswan, P.~K. Jana, and S.~K. Das, ``A survey on mobile charging techniques in wireless rechargeable sensor networks,'' \emph{IEEE Communications Surveys \& Tutorials}, vol.~24, no.~3, pp. 1750--1779, 2022.

\bibitem{5}
W.~Yuan, Z.~Wei, S.~Li, J.~Yuan, and D.~W.~K. Ng, ``Integrated sensing and communication-assisted orthogonal time frequency space transmission for vehicular networks,'' \emph{IEEE Journal of Selected Topics in Signal Processing}, vol.~15, no.~6, pp. 1515--1528, 2021.

\bibitem{6}
Q.~Qi, X.~Chen, A.~Khalili, C.~Zhong, Z.~Zhang, and D.~W.~K. Ng, ``Integrating sensing, computing, and communication in 6g wireless networks: Design and optimization,'' \emph{IEEE Transactions on Communications}, vol.~70, no.~9, pp. 6212--6227, 2022.

\bibitem{7}
T.~Rault, ``Avoiding radiation of on-demand multi-node energy charging with multiple mobile chargers,'' \emph{Computer Communications}, vol. 134, pp. 42--51, 2019.

\bibitem{9}
W.~Xu, W.~Liang, X.~Jia, H.~Kan, Y.~Xu, and X.~Zhang, ``Minimizing the maximum charging delay of multiple mobile chargers under the multi-node energy charging scheme,'' \emph{IEEE transactions on mobile computing}, vol.~20, no.~5, pp. 1846--1861, 2020.

\bibitem{10}
A.~Kaswan, P.~K. Jana, M.~Dash, A.~Kumar, and B.~P. Sinha, ``Dmcp: A distributed mobile charging protocol in wireless rechargeable sensor networks,'' \emph{ACM Transactions on Sensor Networks}, vol.~19, no.~1, pp. 1--29, 2022.

\bibitem{11}
L.~Mo, A.~Kritikakou, and S.~He, ``Energy-aware multiple mobile chargers coordination for wireless rechargeable sensor networks,'' \emph{IEEE internet of things journal}, vol.~6, no.~5, pp. 8202--8214, 2019.

\bibitem{12}
G.~Han, H.~Guan, J.~Wu, S.~Chan, L.~Shu, and W.~Zhang, ``An uneven cluster-based mobile charging algorithm for wireless rechargeable sensor networks,'' \emph{IEEE Systems Journal}, vol.~13, no.~4, pp. 3747--3758, 2018.

\bibitem{13}
A.~Tomar, L.~Muduli, and P.~K. Jana, ``A fuzzy logic-based on-demand charging algorithm for wireless rechargeable sensor networks with multiple chargers,'' \emph{IEEE Transactions on Mobile Computing}, vol.~20, no.~9, pp. 2715--2727, 2020.

\bibitem{14}
Y.~Zhu, K.~Chi, P.~Hu, K.~Mao, and Q.~Shao, ``Velocity control of multiple mobile chargers over moving trajectories in rf energy harvesting wireless sensor networks,'' \emph{IEEE Transactions on Vehicular Technology}, vol.~67, no.~11, pp. 11\,314--11\,318, 2018.

\bibitem{16R}
M.~U.~F. Qaisar, W.~Yuan, P.~Bellavista, G.~Han, R.~S. Zakariyya, and A.~Ahmed, ``Probabilistic on-demand charging scheduling for isac-assisted wrsns with multiple mobile charging vehicles,'' in \emph{GLOBECOM 2023-2023 IEEE Global Communications Conference}.\hskip 1em plus 0.5em minus 0.4em\relax IEEE, 2023, pp. 5895--5900.

\bibitem{15}
M.~U.~F. Qaisar, W.~Yuan, P.~Bellavista, F.~Liu, G.~Han, R.~S. Zakariyya, and A.~Ahmed, ``Poised: Probabilistic on-demand charging scheduling for isac-assisted wrsns with multiple mobile charging vehicles,'' \emph{IEEE Transactions on Mobile Computing}, 2024.

\end{thebibliography}

\vskip -1\baselineskip plus -1fil
\vspace{-11pt}

\begin{IEEEbiography}[{\includegraphics[width=1in,height=1.25in,clip,keepaspectratio]{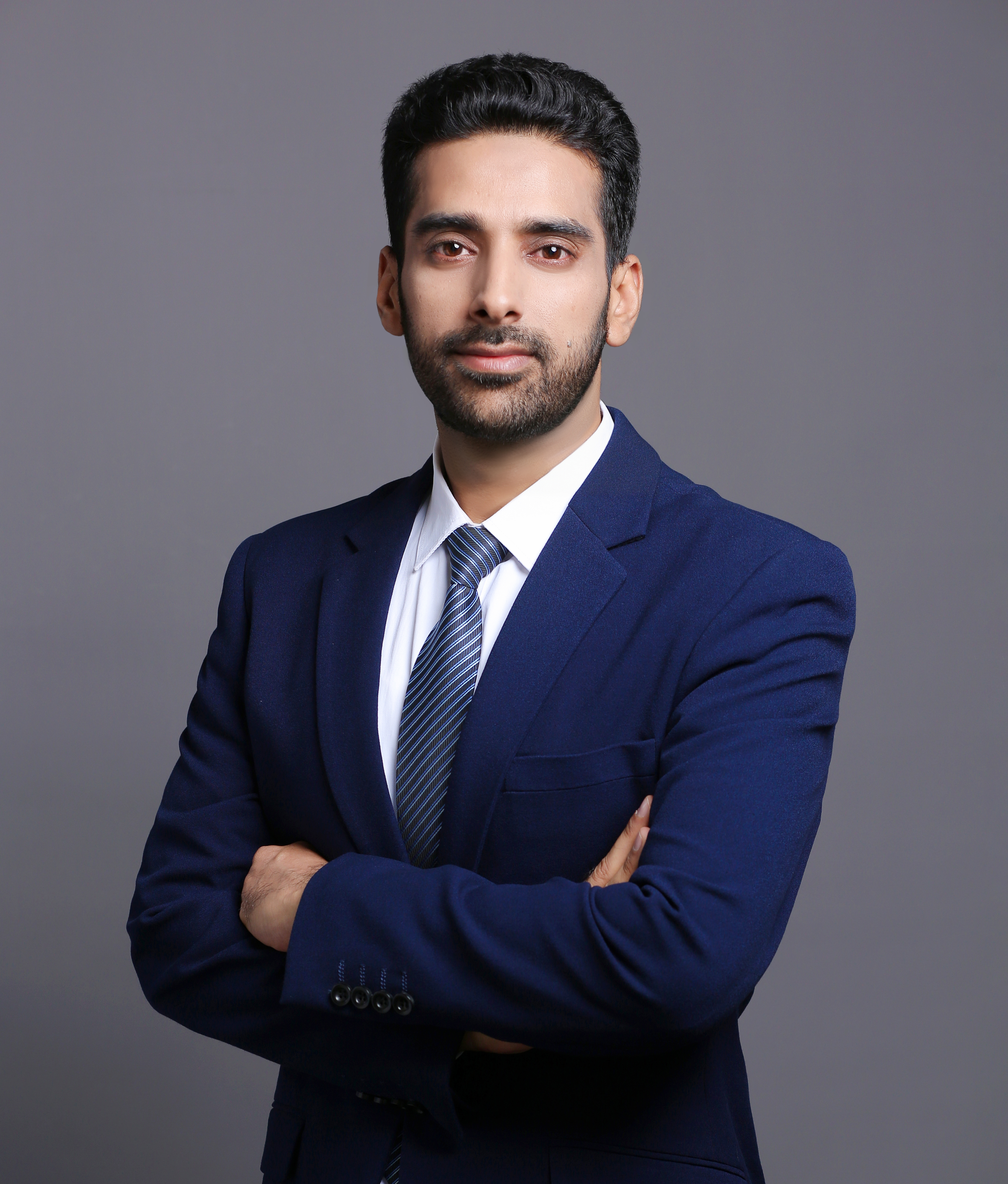}}]{Muhammad Umar Farooq Qaisar}
(Member, IEEE) received his B.S. degree from the International Islamic University Islamabad, Pakistan in 2012 and received his M.S. degree in Computer Science and Technology from the University of Science and Technology of China in 2017. He received his Ph.D. degree in Computer Science and Technology from the University of Science and Technology of China in 2022. He is a postdoctoral fellow in the School of System Design and Intelligent Manufacturing at the Southern University of Science and Technology. His main research interests include IoT, WSN, SDN, VANETs, ISAC, UAVs, and Communication Security.
\end{IEEEbiography}
\vskip -3\baselineskip plus -1fil
\begin{IEEEbiography}[{\includegraphics[width=1in,height=1.25in,clip,keepaspectratio]{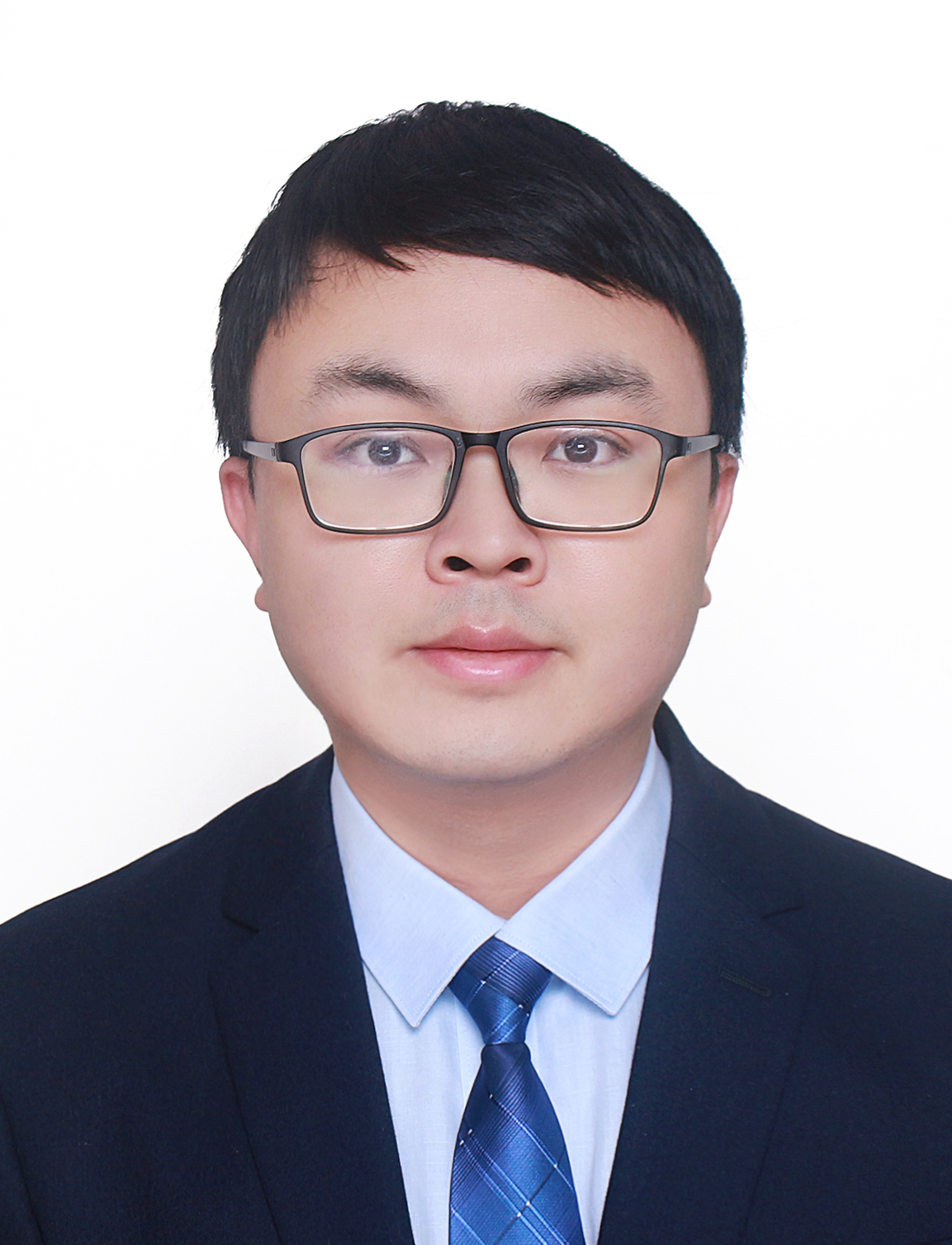}}]{Weijie Yuan}
(Member, IEEE) received the B.E. degree from the Beijing Institute of Technology, China, in 2013, and the Ph.D. degree from the University of Technology Sydney, Australia, in 2019. In 2016, he was a Visiting Ph.D. Student with the Institute of Telecommunications, Vienna University of Technology, Austria. He was a Research Assistant with the University of Sydney, a Visiting Associate Fellow with the University of Wollongong, and a Visiting Fellow with the University of Southampton, from 2017 to 2019. From 2019 to 2021, he was a Research Associate with the University of New South Wales. He is currently an Assistant Professor with the Department of Electrical and Electronic Engineering, Southern University of Science and Technology, Shenzhen, China. He was a recipient of the Best Ph.D. Thesis Award from the Chinese Institute of Electronics and an Exemplary Reviewer from IEEE TCOM/WCL. He currently serves as an Associate Editor for the IEEE Communications Letters, an Associate Editor and an Award Committee Member for the EURASIP Journal on Advances in Signal Processing. He has led the guest editorial teams for three special issues in IEEE Communications Magazine, IEEE Transactions on Green Communications and Networking, and China Communications. He was an Organizer/the Chair of several workshops and special sessions on orthogonal time frequency space (OTFS) and integrated sensing and communication (ISAC) in flagship IEEE and ACM conferences, including IEEE ICC, IEEE GLOBECOM, IEEE/CIC ICCC, IEEE SPAWC, IEEE VTC, IEEE WCNC, IEEE ICASSP, and ACM MobiCom. He is the Founding Chair of the IEEE ComSoc Special Interest Group on Orthogonal Time Frequency Space (OTFS-SIG).
\end{IEEEbiography}

\vskip 0\baselineskip plus -1fil
\vspace{-11 mm}
\begin{IEEEbiography}[{\includegraphics[width=1in,height=1.25in,clip,keepaspectratio]{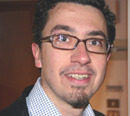}}]{Paolo Bellavista} (Senior Member, IEEE) received the Ph.D. degree in computer science engineering from the University of Bologna, Italy, in 2001. He is currently a Full Professor with the University of Bologna. His research interests include middleware for mobile computing, QoS management in the cloud continuum, infrastructures for big data processing in industrial environments, and performance optimization in wide-scale and latency-sensitive deployment environments. He serves on the Editorial Boards of IEEE Communications Surveys and Tutorials, IEEE Transactions on Network and Service Management, IEEE Transactions on Services Computing, ACM CSUR, ACM TIOT, and PMC (Elsevier). He recently served as the Scientific Coordinator of the H2020 IoTwins Project (https://www.iotwins.eu).
\end{IEEEbiography}
\vskip 0\baselineskip plus -1fil
\vspace{-11 mm}
\begin{IEEEbiography}[{\includegraphics[width=1in,height=1.25in,clip,keepaspectratio]{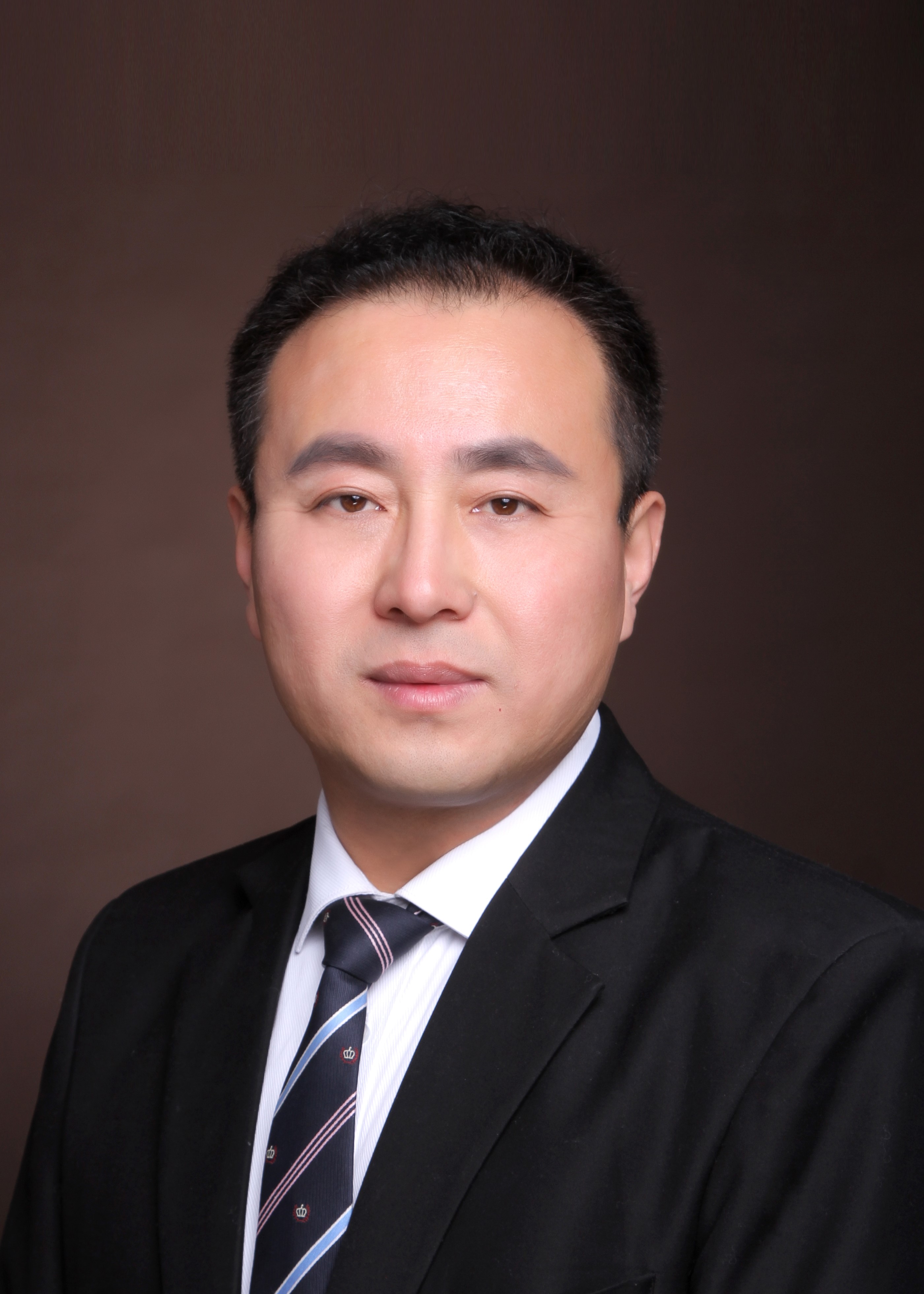}}]{Guangjie Han}
(Fellow, IEEE) is currently a Professor with the Department of Internet of Things Engineering, Hohai University, Changzhou, China. He received his Ph.D. degree from Northeastern University, Shenyang, China, in 2004. In February 2008, he finished his work as a Postdoctoral Researcher with the Department of Computer Science, Chonnam National University, Gwangju, Korea. From October 2010 to October 2011, he was a Visiting Research Scholar with Osaka University, Suita, Japan. From January 2017 to February 2017, he was a Visiting Professor with City University of Hong Kong, China. From July 2017 to July 2020, he was a Distinguished Professor with Dalian University of Technology, China. His current research interests include Internet of Things, Industrial Internet, Machine Learning and Artificial Intelligence, Mobile Computing, Security and Privacy. Dr. Han has over 500 peer-reviewed journal and conference papers, in addition to 160 granted and pending patents. Currently, his H-index is 66 and i10-index is 292 in Google Citation (Google Scholar). The total citation count of his papers raises above 16500+ times. Dr. Han is a Fellow of the UK Institution of Engineering and Technology (FIET). He has served on the Editorial Boards of up to 10 international journals, including the IEEE TII, IEEE TCCN, IEEE TVT, IEEE Systems, etc. He has guest-edited several special issues in IEEE Journals and Magazines, including the IEEE JSAC, IEEE Communications, IEEE Wireless Communications, Computer Networks, etc. Dr. Han has also served as chair of organizing and technical committees in many international conferences. He has been awarded 2020 IEEE Systems Journal Annual Best Paper Award and the 2017-2019 IEEE ACCESS Outstanding Associate Editor Award. He is a Fellow of IEEE.
\end{IEEEbiography}
\vspace{-13 mm}
\vskip 1\baselineskip plus -1fil
\begin{IEEEbiography}[{\includegraphics[width=1in,height=1.25in,clip,keepaspectratio]{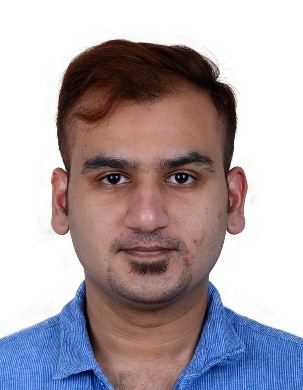}}]{Adeel Ahmed} (Student Member, IEEE) received the BS degree in Telecommunication and Networking from COMSATS University, Pakistan in 2015 and received his M.S. degree in Computer Science and Technology from University of Science and Technology of China in July 2021. Currently, he is pursuing Ph.D. in Computer Science and Technology from University of Science and Technology of China. His research interests include IoT, WBAN, SDN, and SDR. 
\end{IEEEbiography}

\vspace{11pt}

\vfill

\end{document}